\documentclass[aps,showpacs]{revtex4}

\usepackage{amsmath}
\usepackage{graphicx}
\usepackage{amsfonts}
\usepackage{amssymb}
\begin{document}

\title{Diffusion-induced instability and chaos in random oscillator networks}

\author{Hiroya Nakao$^{1}$ and Alexander S. Mikhailov$^{2}$}

\affiliation{$^{1}$Department~of~Physics,~Kyoto~University,~Kyoto~606-8502,~Japan\\
  $^{2}$Abteilung~Physikalische~Chemie,~Fritz-Haber-Institut~der~Max-Planck-Gesellschaft,~Faradayweg~4-6,~14195~Berlin,~Germany}

\begin{abstract}
  We demonstrate that diffusively coupled limit-cycle oscillators on
  random networks can exhibit various complex dynamical patterns.
  Reducing the system to a network analog of the complex
  Ginzburg-Landau equation, we argue that uniform oscillations can be
  linearly unstable with respect to spontaneous phase modulations due
  to diffusional coupling - the effect corresponding to the
  Benjamin-Feir instability in continuous media.  Numerical
  investigations under this instability in random scale-free networks
  reveal a wealth of complex dynamical regimes, including partial
  amplitude death, clustering, and chaos. A dynamic mean-field theory
  explaining different kinds of nonlinear dynamics is constructed.
\end{abstract}

\date{October 10, 2008}

%%
%% 05.45.-a Nonlinear dynamics and chaos
%% 05.45.Xt Synchronization; coupled oscillators
%% 89.75.Fb Structures and organization in complex systems

\maketitle

Phase oscillators coupled through various network structures have been
extensively analyzed as a prototype model of network
dynamics~\cite{Strogatz,Moreno,Ichinomiya,Arenas}.
In most studies, complete synchronization of all oscillators has been
the main focus, though possibilities of more complex dynamics have
also been reported~\cite{Ko,Gil}.
Coupled phase oscillators are obtained from general coupled
limit-cycle oscillators by eliminating amplitude degrees of freedom in
the weak coupling limit~\cite{Kuramoto}.
When the coupling is not weak, such phase reduction breaks down and
much richer dynamics can be expected.

In this paper, we analyze complex dynamics exhibited by diffusively
coupled limit-cycle oscillators on random networks.
In continuous media, sufficiently large difference in diffusion
constants of oscillating components (e.g. chemical species) can
destabilize uniform oscillations and lead to diffusion-induced
spatiotemporal chaotic regimes~\cite{Kuramoto}, such as those
experimentally observed in surface chemical reactions~\cite{Kim}.
We argue that diffusional mobility of the components can also lead to
the instability and complex dynamics on networks.

Rather than treating a specific model of limit-cycle oscillators, we
focus on a network version of the complex Ginzburg-Landau (CGL)
equation derived from a general model of diffusively coupled
limit-cycle oscillators near the supercritical Hopf bifurcation.
Our linear stability analysis based on Laplacian eigenvectors of the
network generally shows that the uniformly oscillating solution can
become unstable when the analog of the Benjamin-Feir (BF) condition is
satisfied.
Numerical simulations on random scale-free networks under this
condition reveal different kinds of complex dynamical regime.
To explain them, an approximate mean-field theory is constructed.

We consider a system of diffusively coupled identical limit-cycle
oscillators on random networks consisting of $N$ nodes described by
\begin{align}
  \label{ODE}
  \dot{\bf X}_j(t) = {\bf F}({\bf X}_j) + {\bf D} \sum_{k=1}^{N}
  L_{jk} {\bf X}_k.
\end{align}
Here, ${\bf X}_j(t)$ represents the state of the oscillator on node
$j$ ($j=1, \cdots, N$), ${\bf F}({\bf X})$ specifies the intrinsic
dynamics of an oscillator, and the last term takes into account
diffusive coupling on the network, where ${\bf D}$ is a diffusion
matrix and $L_{jk}$ is a Laplacian matrix of the network.
The network is defined by a symmetric adjacency matrix $A_{jk}$, whose
components are $1$, if the nodes $j$ and $k$ are connected, and $0$
otherwise.
The Laplacian matrix is given by $L_{jk} = A_{jk} - k_j \delta_{jk}$
with $k_j = \sum_{k=1}^{N} A_{jk}$ representing the degree (number of
connections) of node $j$.
We assume that each oscillator has a stable limit-cycle solution ${\bf
  X}^0(t)$ in absence of diffusion.
A uniformly oscillating solution of the system, ${\bf X}_j(t) \equiv
{\bf X}^0(t)$ for $\forall j$, always satisfies Eq.~(\ref{ODE})
because $\sum_{k=1}^{N} L_{jk} = 0$ holds, but diffusion may
destabilize this solution.

We assume that each oscillator is slightly above the supercritical
Hopf bifurcation point and consider a situation where the effect of
diffusion is also comparably small.
Then, using the standard weakly nonlinear analysis~\cite{Kuramoto}, we
can reduce Eq.~(\ref{ODE}) to a network version of the CGL (or
Kuramoto-Tsuzuki) equation,
\begin{align}
  \label{CGL}
  \dot{W}_j(t) = (1 + i c_0) W_j - (1 + i c_2) \left| W_j \right|^2
  W_j + K (1 + i c_1) \sum_{k=1}^{N} L_{jk} W_k.
\end{align}
Here, $W_j(t)$ represents the complex oscillation amplitude of $j$-th
oscillator such that ${\bf X}_{j}(t) - {\bf X}^{(S)} \propto W_{j}(t)
\exp(i \omega_0 t) {\bf U} + c.c.$ where ${\bf X}^{(S)}$ is the
unstable fixed point, $\omega_0$ is the Hopf frequency, and ${\bf U}$
is the complex critical eigenvector of the Jacobian matrix of ${\bf
  F}({\bf X})$ at ${\bf X}^{(S)}$.
Real parameters $c_0$, $c_1$, $c_2$, and positive coupling strength
$K$ can be determined when ${\bf F}({\bf X})$ and ${\bf
  D}$ are explicitly given.
Note that if the diffusion constants of all components are equal, i.e.
${\bf D} = D {\bf I}$ where ${\bf I}$ is the identity matrix, we have
$c_1 = 0$.
Equation~(\ref{CGL}) has a uniformly oscillating solution, $W_{j}(t)
\equiv W^{0}(t) := \exp[ i ( c_{0} - c_{2} ) t ]$ for $\forall j$.

When $K$ is small enough, each oscillator state is always near the
unperturbed limit cycle, so that Eq.~(\ref{CGL}) can further be
reduced to coupled phase oscillators of the form $\dot{\phi}_j(t) =
\omega - C \sum_{k=1}^{N} L_{jk} \sin( \phi_{j} - \phi_{k} + \gamma)$,
where $\phi_{j}$ is the phase of the oscillator $j$, $\omega = c_{0} -
c_{2}$ is the frequency, $C$ is the rescaled coupling strength, and
the coupling phase shift $\gamma$ satisfies $\cos \gamma = ( 1 + c_{1}
c_{2} ) / \sqrt{ (1 + c_{1}^2 ) (1 + c_{2}^2 ) }$.
Recently, it has been shown that this network phase model exhibits
coexistence of drifting and phase-locked oscillators with stationary
phase gradients~\cite{Ko}.
In the following, we focus on the case with stronger coupling.

Let us analyze linear stability of the uniform solution.
Plugging weakly perturbed solution $W_{j}(t) = W^{0}(t) \{ 1 +
\rho_{j}(t) \} \exp[ i \theta_{j}(t) ]$ into Eq.~(\ref{CGL}) with
$\rho_{j}(t)$ and $\theta_{j}(t)$ being amplitude and phase
perturbations, respectively, we obtain the following linearized
equations:
\begin{align}
  \dot{\rho_{j}}(t) &= - 2 \rho_{j} + K \sum_{k=1}^{N} L_{jk} (
  \rho_{k} - c_{1} \theta_{k} ), \cr 
  \dot{\theta_{j}}(t) &= - 2 c_{2} \rho_{j} + K \sum_{k=1}^{N} L_{jk}
  ( c_{1} \rho_{k} + \theta_{k} ).
\end{align}
To proceed, we introduce Laplacian eigenvalues $\Lambda^{(\alpha)}$
and eigenvectors ${\boldsymbol \phi}^{(\alpha)} = (
\phi_{1}^{(\alpha)}, \cdots, \phi_{N}^{(\alpha)} )$ of the Laplacian
matrix $L_{jk}$ satisfying $\sum_{k=1}^{N} L_{jk}\phi_{k}^{(\alpha)} =
\Lambda^{(\alpha)} \phi_{j}^{(\alpha)}$ for $\alpha= 1, \cdots, N$.
All eigenvalues are real and non-positive, and the eigenvectors are
mutually orthogonal.
We expand the perturbations as $(\rho_{j}, \theta_{j}) =
\sum_{\alpha=1}^{N} (\rho^{(\alpha)}, \theta^{(\alpha)})
\phi_{j}^{(\alpha)} \exp \left( \lambda^{(\alpha)} t \right)$ where
$\rho^{(\alpha)}$ and $\theta^{(\alpha)}$ are expansion coefficients
and $\lambda^{(\alpha)}$ is the complex growth rate of $\alpha$-th
eigenmode.
Then, a characteristic equation $\left\{ \lambda^{(\alpha)}
\right\}^{2} + 2 \{ 1 - K \Lambda^{(\alpha)} \} \lambda^{(\alpha)} - 2
(1 + c_{1} c_{2}) K \Lambda^{(\alpha)} + (1 + c_{1}^{2}) \left\{ K
  \Lambda^{(\alpha)} \right\}^{2} = 0$ is obtained for each eigenmode,
which yields
\begin{align}
  \label{Eigen}
  \lambda_{\pm}^{(\alpha)} &= - 1 + K \Lambda^{(\alpha)} \pm \sqrt{ 1
    + 2 c_{1} c_{2} \left\{ K \Lambda^{(\alpha)} \right\} - c_{1}^{2}
    \left\{ K \Lambda^{(\alpha)} \right\}^{2}}.
\end{align}
When $\mbox{Re}\ \lambda_{\pm}^{(\alpha)} > 0$ for some $\alpha$, the
$\alpha$-th eigenmode is unstable.

\begin{figure}[t]
  \begin{center}
    \includegraphics[width=1.0\hsize]{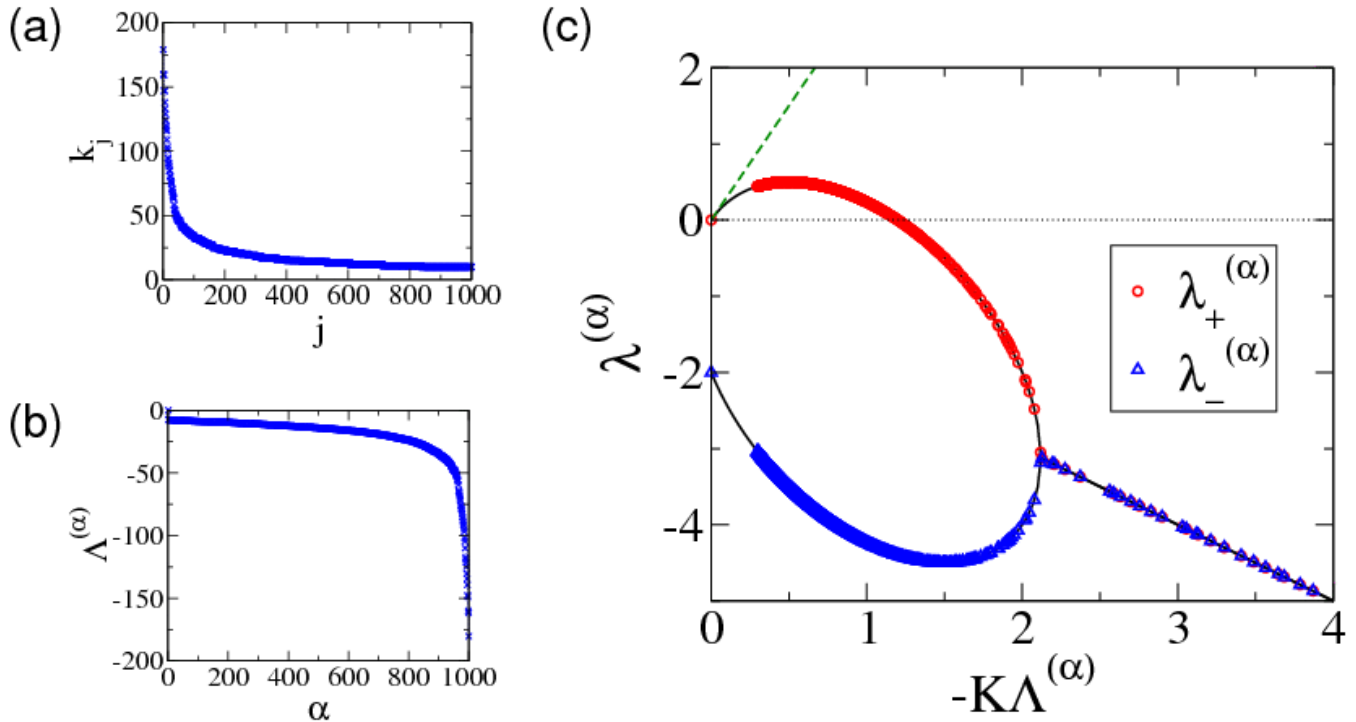}
    \caption{(Color online) (a) Degree $k_j$ and (b) Laplacian
      eigenvalue $\Lambda^{(\alpha)}$ of the scale-free network used
      in numerical simulations.  (c) Linear growth rates of
      perturbations $\mbox{Re}\ \lambda_{\pm}^{(\alpha)}$ plotted as
      functions of $- K \Lambda^{(\alpha)}$.  Dashed straight line
      shows the slope $- (1+c_1 c_2)$ of the upper branch at the
      origin.}
    \label{Fig1}
  \end{center}
\end{figure}

By expanding the upper branch $\lambda_{+}^{(\alpha)}$ of
Eq.~(\ref{Eigen}) for small $K \Lambda^{(\alpha)}$, we obtain
$\lambda_{+}^{(\alpha)} = (1 + c_{1} c_{2} ) K \Lambda^{(\alpha)} + O(
\left\{ K \Lambda^{(\alpha)} \right\}^2 )$.
Therefore, $\mbox{Re}\ \lambda_{+}^{(\alpha)}$ can be positive when
the condition $1 + c_{1} c_{2} < 0$ is satisfied (note that
$\Lambda^{(\alpha)} \leq 0$).
This is the same as the BF condition for instability of the uniform
solution of the CGL equation in continuous media~\cite{Kuramoto},
which also applies to globally coupled and non-locally coupled CGL
oscillators~\cite{GCGL,NCGL}.
Note that the BF condition cannot be satisfied for $c_{1} = 0$ and
therefore a sufficiently large difference in diffusion constants of
the components is necessary.
For the instability to actually occur, the discrete Laplacian
eigenvalues should exist near the peak of the upper curve given by
Eq.~(\ref{Eigen}).
As we already know for other coupling
schemes~\cite{Kuramoto,GCGL,NCGL}, Eq.~(\ref{CGL}) is expected to
exhibit strongly nonlinear behavior once the uniform solution becomes
unstable.

As an example of random networks, we use random scale-free networks of
size $N=1000$ and mean degrees $\langle k \rangle = 20$ generated by
the B\'arabasi-Albert preferential attachment rule~\cite{Barabasi}.
We fix the parameters $c_{1}=-2$ and $c_{2}=2$ ($c_{0}$ can be set to
$0$ without loss of generality), and vary the coupling strength $K$.
Numerical results shown below are for one particular realization of
the random network, but similar behavior was observed for other
network realizations as well.

Figure~\ref{Fig1}(a) displays the degree $k_{j}$ of each node vs. the
node index $j$, where the node indices $\{ j \}$ are sorted in
decreasing order of their degrees $\{ k_{j} \}$ so that inequalities
$k_{1} \geq k_{2} \cdots \geq k_{N}$ hold.  We use this ordering as a
useful way to visualize the complex dynamics on the network throughout
our analysis.
Figure~\ref{Fig1}(b) shows the Laplacian eigenvalues
$\Lambda^{(\alpha)}$ of the same network.  The eigenvalue indices
$\{\alpha\}$ are also sorted in decreasing order of the eigenvalues
such that $0 = \Lambda ^{(1)} > \Lambda^{(2)} > \cdots >
\Lambda^{(N)}$ hold.

Figure~\ref{Fig1}(c) plots growth rates of the perturbations
$\mbox{Re}\ \lambda_{\pm}^{(\alpha)}$ obtained by the linear stability
analysis as functions of $- K \Lambda^{(\alpha)}$ at $K = 0.04$.  The
actual growth rates are distributed discretely on the curves given by
Eq.~(\ref{Eigen}).
We can see that the growth rates on the upper branch $\mbox{Re}\
\lambda_{+}^{(\alpha)}$ can become positive when the coupling strength
$K$ is in an appropriate range, indicating that the uniform solution
can undergo a diffusion-induced instability.

\begin{figure}[t]
\begin{center}
  \includegraphics[width=1.0\hsize]{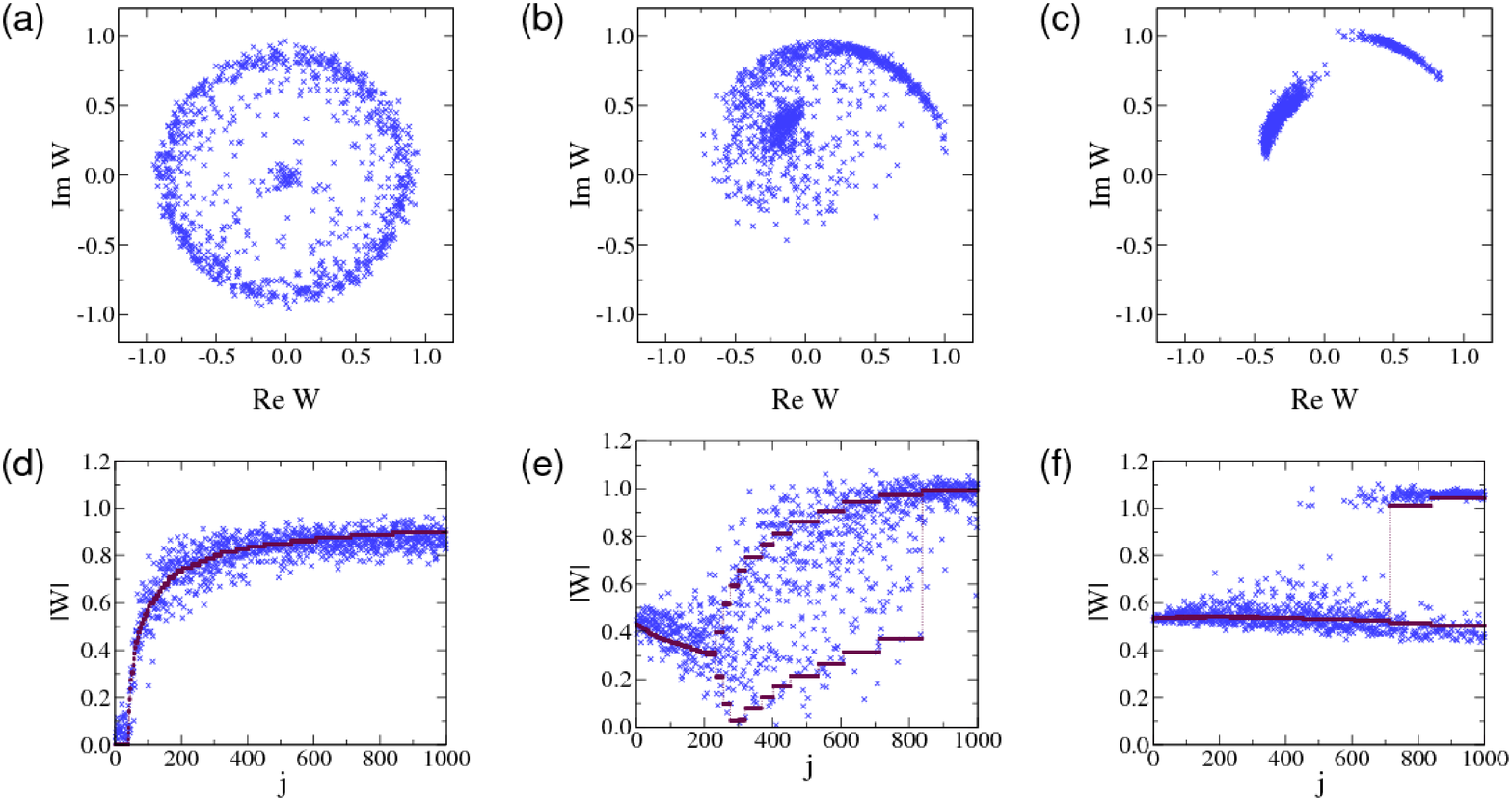}
  \caption{(Color online) (a-c): Snapshots of the complex amplitude
    $W_j$ on the complex plane.  (d-f): Snapshots of the amplitude
    $|W_j|$ vs. the node index $j$.  Coupling strength is $K=0.02$ for
    (a,d), $K=0.04$ for (b,e), and $K=0.08$ for (c,f).  
    Node indices are sorted in decreasing
    order of their degrees $\{ k_{j} \}$ so that inequalities
    $k_{1} \geq k_{2} \cdots \geq k_{N}$ hold.
    Solid curves in (d-f) are predictions of the mean-field theory.}
  \label{Fig2}
\end{center}
\end{figure}

To investigate nonlinear dynamics after the instability, we have
performed numerical simulations of Eq.~(\ref{CGL}) with slightly
perturbed uniform solutions as initial conditions.
When $K$ is very small ($K < 0.001$), no oscillator deviates largely
from the unperturbed limit-cycle orbit $W^{(0)}(t)$, so that the
reduced phase model is valid.  The coupling phase shift is given by
$\gamma = \arccos(-3/5) \simeq -2.21$, which is repulsive because
$|\gamma| > \pi/2$~\cite{Kuramoto}.  Therefore, the oscillators do not
synchronize but rotate incoherently.
When $K$ is very large ($K > 0.165$), there exist no discrete growth
rates on the positive part of the upper curve of Fig.~\ref{Fig1}(c),
so that the uniform solution remains stable even if the BF condition
is satisfied.

Between these limits, we have found three characteristic steady
dynamical regimes as shown in Fig.~\ref{Fig2}, where snapshots of the
amplitude profile $|W_{j}|$ and the distribution of $W_{j}$ on the
complex plane are displayed for three values of the coupling strength,
$K=0.02$, $K=0.04$, and $K=0.08$.

(i) {\em Partial amplitude death} [Figs.~\ref{Fig2}(a),(d)].  When
$0.006 < K < 0.028$, a group of oscillators with small node indices
(i.e. with large degrees) stops rotation and stays near the origin of
the complex plane while other oscillators are rotating around circular
orbits incoherently, with a rather sharp but smooth transition between
the two groups.

(ii) {\em Chaos} [Figs.~\ref{Fig2}(b),(e)]. When $0.028 < K < 0.078$,
the oscillators are roughly separated into three groups.
In the first group, oscillators take approximately constant amplitudes
near $0.5$, which corresponds to the central cluster on the complex
plane.
Amplitudes of oscillators in the second group are strongly scattered
and evolve chaotically, but their envelope still forms smooth curves.
This group corresponds to the intermediate scattered oscillator states
on the complex plane.
The oscillators in the last group again take constant amplitudes near
$1$, which correspond to the oscillator states elongated along the
unit circle on the complex plane.
The largest Lyapunov exponent of the system is positive in this
regime.

(iii) {\em Clustering} [Figs.~\ref{Fig2}(c),(f)]. When $0.078 < K <
0.164$, phase relations among the oscillators are frozen and the whole
system exhibits a rigid constant rotation.
For relatively small values of $K$ ($K < 0.12$), the oscillators
with small degrees split into two groups with two distinct amplitudes,
i.e. they exhibit a 2-cluster state.
As $K$ becomes larger, the two clusters gradually approach each other
and, at relatively large $K$ ($K > 0.14$), the two clusters merge to a
single cluster but still with phase scattering.

Transitions between the above dynamical regimes occur abruptly and are
clearly detectable, whereas the change in the dynamics within each
regime, e.g. transformation from 2-cluster to 1-cluster states, occurs
gradually with $K$.

\begin{figure}[htbp]
  \begin{center}
    \includegraphics[width=0.7\hsize]{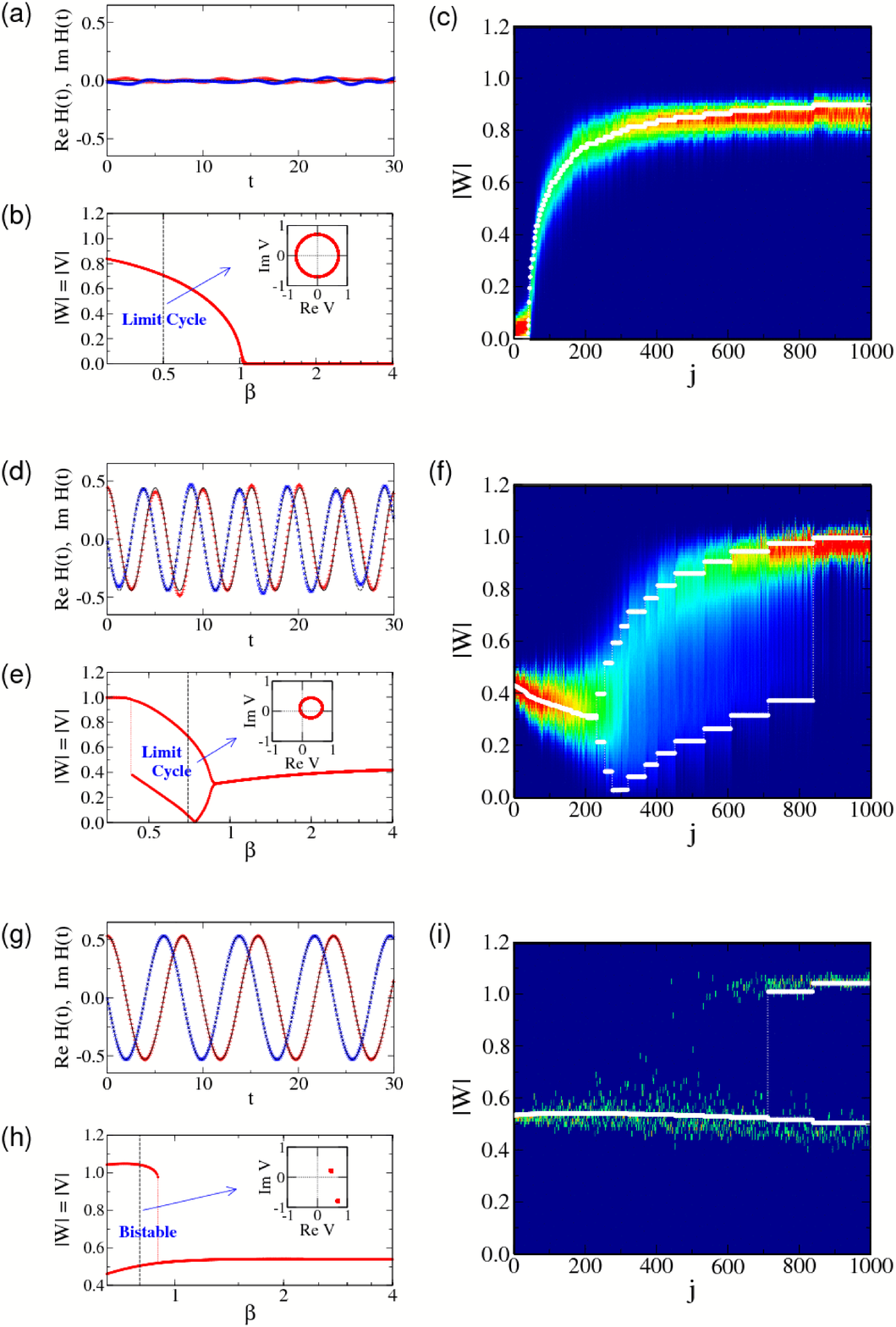}
    \caption{(Color online) (a), (d), (g): Evolution of real and
      imaginary parts of the global mean field $H(t)$ and fitting by
      $B \exp( i \Omega t )$. (b), (e), (h): Bifurcation diagrams of
      the sinusoidally-driven oscillator.  Insets show limit cycle
      orbits or fixed points of $V(t)$ at the parameter values
      indicated by broken vertical lines.  (c), (f), (i): Probability
      density functions of $|W_j|$ compared with the mean-field
      approximation.  Solid curves represent the maximal and minimal
      values of the complex amplitude $|W_j|$.  Parameters are
      $K=0.02$, $B=0$ (a,b,c), $K=0.04$, $B=0.442$, $\Omega=-1.25$
      (d,e,f), and $K=0.08$, $B=0.532$, $\Omega=-0.796$ (g,h,i).  }
      \label{Fig3}
  \end{center}
\end{figure}

To explain the observed dynamical patterns, we employ the mean-field
approximation, valid for large random networks with strong diffusive
mixing.  It has been used in analyzing network-based epidemics
spreading models~\cite{Pastor}, coupled phase
oscillators~\cite{Ichinomiya,Ko}, and also network Turing
patterns~\cite{Turing}.
A crucial point here is that we consider not only static but also
dynamic mean fields that oscillate periodically with time.

Introducing a complex local field $h_{j}(t) = \sum_{k=1}^{N} A_{jk}
W_{k}(t)$, the diffusion term in Eq.~(\ref{CGL}) can be written as
$\sum_{k=1}^{N} L_{jk} W_{k} = h_{j}(t) - k_{j} W_{j}$.
We approximate this local field as
\begin{align}
  \label{MF}
  h_{j}(t) \simeq k_{j} H(t), \;\;\; H(t) = \sum_{j=1}^{N}
  \frac{k_{j}}{k_{total}} W_{j}(t),
\end{align}
where $k_{total} = \sum_{j=1}^{N} k_{j}$ and $H(t)$ is a
degree-weighted global mean field over the
network~\cite{Pastor,Ichinomiya,Turing}.
Thus, we ignore detailed connections of the network and retain only
the degrees.
Equation~(\ref{CGL}) is then approximated as
\begin{align}
  \label{SLMF}
  \dot{W}_j(t) = (1 + i c_0) W_j - (1 + i c_2) \left| W_j \right|^2
  W_j + k_{j} K (1 + i c_1) \left\{ H(t) - W_j \right\},
\end{align}
which describes independent CGL oscillators coupled to a global mean
field $H(t)$.  The effective coupling strength of each oscillator to
$H(t)$ is given by $k_{j} K$, and thus depends on the node degree
$k_j$.

In Figs.~\ref{Fig3}(a), (d), and (g), time sequences of the global
mean field $H(t)$ obtained numerically for the three cases in
Fig.~\ref{Fig2} are shown.
$H(t)$ almost vanishes at $K=0.02$, whereas it oscillates sinusoidally
at $K=0.04$ and $K=0.08$.
We can thus approximate $H(t)$ in these regimes as $ H(t) = B \exp
\left( i \Omega t \right) $, where $B$ and $\Omega$ denote amplitude
and frequency of the periodic sinusoidal oscillation, which reasonably
fit the numerical data as shown in the figures.
Similar sinusoidal-field approximation has been used in the analysis
of collective dynamics of globally coupled CGL
oscillators~\cite{GCGL}, but degree inhomogeneity in networks
essentially changes the results.
Precisely speaking, in the chaotic regime, $H(t)$ is only
approximately sinusoidal and can be more complex, e.g. quasiperiodic
for some other values of $K$ (as also known in the case of global
coupling~\cite{GCGL}), but we focus on the simplest sinusoidal case
here.  In the clustering regime, $H(t)$ is always strictly sinusoidal.

Moving to a rotating frame by introducing $W(t) = V(t) \exp \left( i
  \Omega t \right)$, we obtain an autonomous equation for $V(t)$ as
$\dot{V}(t) = [ 1 + i (c_0 - \Omega) ] V - (1 + i c_2) \left| V
\right|^2 V + \beta (1 + i c_1) \left( B - V \right)$.
Here we dropped the index $j$, because all oscillators obey the same
dynamics, and defined $\beta = \beta(j) = k_{j} K$, which plays the
role of a bifurcation parameter.  The dependence of the oscillator
dynamics on the node index $j$ enters only through $\beta$.

Figures~\ref{Fig3}(b), (e), and (h) display the bifurcation diagrams
of the above equation as functions of the control parameter $\beta$,
where the maximal and the minimal values of $|W| = |V|$ are plotted
using $B$ and $\Omega$ estimated numerically in Figs.~\ref{Fig3}(a),
(d), and (g).
Depending on the values of $B$, $\Omega$, and $\beta$, the equation
exhibits a symmetric limit cycle, an asymmetric limit cycle, and one
or two fixed points~
\footnote{In~\cite{GCGL}, it is reported that
the asymmetric limit cycle can coexist with the fixed points
in a certain parameter region, where the transition with hysteresis
occurs via a saddle-node bifurcation of the fixed points followed by a
homoclinic or Hopf bifurcation of the
limit cycle.  In the case of Figs.~\ref{Fig3}(e), the
transition near $\beta = 0.435$ occurs by a SNIPER (saddle-node infinite-period on limit cycle) bifurcation without hysteresis, and coexistence of fixed points with the asymmetric limit cycle does not take place.  Note that coexistence of
multiple attractors in the dynamics of individual oscillators, which
actually occurs in the $2$-cluster state (Fig.~\ref{Fig3}(h)),
does not affect our argument as long as the global mean field is
approximately sinusoidal.}.

Now, using the relation between the bifurcation parameter and the node
degree, $\beta = \beta(j) = K k_j$, we can map the bifurcation
diagrams onto actual amplitude patterns in the network.
The solid curves in Figs.~\ref{Fig2}(d), (e), and (f) are the maximal
and minimal values of $|W_{j}| = |V_{j}|$, which fit the envelopes of
the oscillator dynamics reasonably well.
Figure~\ref{Fig3}(c), (f), and (i) compare the numerical probability
density functions of the amplitude with these curves, showing good
agreement.
In particular, the condition for an oscillator to fall in the
amplitude death state in regime (i) can be obtained analytically by
linear stability analysis of the fixed point $V_{j} = 0$ with $B=0$.
This yields $k_j > 1 / K$, which also agrees well with the numerical
data.
Thus, the complex network dynamics in our model can be well understood
through the mean-field approximation~
\footnote{Though we do not give details in the present paper,
we can further develop a self-consistency analysis for the
sinusoidal global mean field $H(t)$ to determine $B$ and $\Omega$
from the condition that the $H(t)$ imposed to Eq.~(\ref{SLMF})
coincides with the $H(t)$ resulting from Eq.~(\ref{MF}).  This gives good agreement with direct numerical results for $0 \leq K < 0.05$,
where $H(t)$ vanishes or oscillates sinusoidally and each individual
oscillator has a single attractor.  For larger values of $K$ where $H(t)$
is not sinusoidal or some of the individual oscillators have multiple
attractors (e.g. bistable fixed points), such a simple sinusoidal
self-consistency analysis fails.  See Chabanol {\it et al.}~\cite{GCGL} for elaborate analysis of the globally coupled CGL oscillators.}.

Summarizing, we have investigated diffusion-induced instability and
resulting complex dynamics exhibited by limit-cycle oscillators on
random networks.
Under the mean-field approximation, the observed inhomogeneous
dynamical patterns can be interpreted as a mixture of various limit
cycles and fixed points, which is reminiscent of the ``chimera''
states found in nonlocally coupled oscillators~\cite{Chimera,Ko}.  In
the present case, however, the degree inhomogeneity of the network
essentially determines the dynamics of each oscillator.

Dynamical systems coupled through various networks are ubiquitous
structures in the real world, ranging from neuronal circuits in the
brain to various engineering problems, such as sensor networks and
power grids (see \cite{Strogatz,Arenas}).
The fact that complex dynamical patterns can spontaneously emerge in
random oscillator networks may be of fundamental importance in
understanding the behavior and functions of such systems.

Financial support of the Volkswagen Foundation (Germany) and the MEXT
(Japan, Kakenhi 19762053) is gratefully acknowledged.

\end{document}